\documentclass[prd,aps,showpacs]{revtex4}
\usepackage{amsmath,amssymb}
\usepackage{epsf}
\newcommand{\be}{\begin{equation}}
\newcommand{\ee}{\end{equation}}
\newcommand{\bea}{\begin{eqnarray}}
\newcommand{\eea}{\end{eqnarray}}
\begin{document}
%%%%%%%%%%%%%%%%%%%%%%%%%%%%%%%%%%%%%%%%%%%%%%%%%%%%%%%%%%%%%%%%%%%
\begin{flushright}
        \hfill{CERN-PH-TH/2007-045} \\
\end{flushright}

\vspace{5pt}
\title{\bf Non-Abelian vortices in Chern-Simons theories
and their induced effective theory}
\author{L.G.~Aldrovandi$^{a,b}$ and
F.A.~Schaposnik$^{b,c}$\footnote{F.A.S. is associated with CICBA.}}

\affiliation{{$^a$}CERN, Theory Division\\ CH-1211, Geneva 23,
Switzerland. \\$^b$Departamento\ de F\'{\i}sica, Universidad
Nacional de La Plata,
      C.C. 67, (1900) La Plata, Argentina.\\
      $^c$CEFIMAS, Santa Fe 1145, C1059ABF - Buenos Aires,
      Argentina
      }
\date{\today}

%===================================================================
\begin{abstract}
Non-Abelian vortices for a supersymmetric ${\cal N}=2$
Chern-Simons-Higgs theory are explicitly constructed. We introduce
$N$ Higgs fields in the fundamental representation of the $U(N)$ gauge group
in order to have a color-flavor $SU(N)$ group remaining unbroken in the
asymmetric phase. Bogomol'nyi-like first order equations are found and
rotationally symmetric solutions are proposed. These solutions are
shown to be truly non-Abelian by parameterizing them in terms of
orientational collective coordinates. The low energy effective
action for the orientational moduli results to be the
one-dimensional supersymmetric ${\cal N}=2$ ${\bf CP}^{N-1}$ model. We analyze the
quantum mechanics of this effective theory in the $N=2$ case.
\end{abstract}
\maketitle

\section{Introduction}

Planar physics exhibits many interesting features, both
theoretically and experimentally. This is due in part to the fact
that the behavior of matter and gauge fields in $d=2$ spatial
dimensions can differ in radical ways from the standard behavior we
are used to in $d=3$. A clear example of this is the possibility of
considering Chern-Simons terms
\cite{Deser:1982vy},\cite{Deser:1981wh} whose dynamics is completely
different from that of Yang-Mills theories. Chern-Simons gauge
theories have been extensively studied in the past, showing several
unusual features. For instance, the Abelian theories contain
particle excitations carrying fractional spin and statistics, an
aspect which has been widely used in many fields in physics, such as
the fractional spin in Quantum field theory
\cite{Frohlich:1988qh}-\cite{3}, and the quantum Hall effect in
condensed matter physics \cite{Zhang:1988wy}-\cite{mur}. It should
be also noticed that the high temperature limit of a four
dimensional field theory can be described by a three dimensional
theory where a Chern-Simons term enters naturally. For a nice
review, see Ref.\cite{Dunne:1998qy}.

Furthermore, Abelian Chern-Simons-Higgs systems become self-dual
when the Higgs potential takes a special sixth order form so that
the energy functional of the systems has a Bogomol'nyi bound
\cite{Bogomolny:1975de}, which is saturated by the solutions of the
self-duality equations \cite{hong}. The sixth-order potential has
both a symmetric and an asymmetric vacuum. Solutions approaching the
asymmetric vacuum at infinity describe topologically stable
vortices, whereas solutions which approach the symmetric phase at
infinity are non-topological \cite{Jackiw:1990pr}. The same
self-dual structure has also been found in $SU(N)$ non-Abelian
Chern-Simons-Higgs systems when the Higgs field is in the
fundamental representation \cite{Lee:1990ep}. The vortices in this
case reduce to the Abelian ones, but only stable non-topological
solutions are present in both the symmetric and asymmetric phases.

Previously, Yang-Mills-Chern-Simons-Higgs theories with the Higgs field in the
adjoint representation were investigated for the gauge groups
$SU(2)$ \cite{deVega:1986eu} and $SU(N)$ \cite{deVega:1986hm}.
By including $N$ Higgs
multiplets in the $SU(N)$ theory (to ensure the maximal breaking
of the gauge symmetry), topologically stable non-selfdual vortex solutions were
found. These were the first examples known of the so-called
${\mathbb Z}_N$ strings, i.e. strings in non-Abelian theories
associated with the center of the $SU(N)$ gauge group. Since
the models considered contained only a fourth-order potential, these solutions were
not found to be self-dual. Self-dual vortex solutions to $SU(2)$ Chern-Simons-Higgs
theory with a sixth-order potential were constructed in \cite{manias}. However, in
these constructions the gauge flux is always directed along a
fixed vector in the Cartan subalgebra of $SU(N)$, and no moduli
which would make the flux orientation a dynamical variable were
found. Therefore, these strings are in this sense Abelian.

Recently, significant progress has been achieved in obtaining
genuinely non-Abelian strings in four-dimensional Yang-Mills
theories, both ${\cal N} = 2$ supersymmetric and
non-supersymmetric \cite{Hanany:2003hp}-\cite{Gorsky:2004ad}. The
bosonic sector of these theories consists basically in a $U(N)$
gauge theory coupled to $N$ Higgs field in the fundamental
representation. The resulting models support non-Abelian magnetic
flux tubes and non-Abelian confined magnetic monopoles at weak
coupling. They are characterized by the
presence of orientational moduli associated with the rotation of
their color flux in the non-Abelian gauge group $SU(N)$.

It is the purpose of this work to look for non-Abelian string
solutions when gauge field dynamics is solely governed by a Chern-Simons
action and the symmetry breaking potential is
sixth-order in order to ensure self-duality and supersymmetric extension.
 It should be stressed that only in the presence of the
Chern-Simons term electrically charged vortices with finite energy
(per unit length) do exist \cite{Julia:1975ff},
\cite{Deser:1982vy},\cite{Deser:1981wh} and hence, the model
we are interested in could show novel aspects of non-Abelian
string-like configurations. Besides, the existence of some kind of
anyons with new internal degrees of freedom, as those given by an
orientational moduli space, could result very useful with respect
to their application to condensed matter problems \cite{Dunne:1998qy}.

The paper is organized as follows: we start by considering in
section 2 a supersymmetric ${\cal N}=2$ Chern-Simons theory with
$U(N)$ gauge group coupled to $N$ scalar multiplets. We obtain for
this model the pattern of symmetry breaking and show that in the
asymmetric vacuum, the theory lies in the so-called colour-flavor locked
phase. In section 3, we begin obtaining the self-duality equations
through the usual Bogomol'nyi trick. Next, we find
non-Abelian vortices in this model by considering a rotationally
symmetric ansatz in the topological sector. These strings are 1/2
BPS saturated and are shown to have a non-Abelian character due to
the existence of a set of orientational collective coordinates.
Section 4 is devoted to find the low-energy world-line theory
describing moduli dynamics which turns out to be the
one-dimensional supersymmetric ${\cal N}=2$ ${\bf CP}^{N-1}$ model. The quantum
mechanics of the effective theory is analyzed in section 5 for the
case $N=2$. A summary and discussion of our results are presented
in section 6.

\section{The Theory}

Throughout this paper we shall consider the ${\cal N}=2$
supersymmetric extension of the  $d=2+1$ dimensional
non-Abelian Chern-Simons-Higgs. The field content of the theory is given by a
$U(N)$ vector multiplet consisting in the gauge field $A_\mu $,
where $\mu,\nu,...=0,1,2$ are Lorentz indices, coupled to $N$
scalar multiplets, each of which contain a complex scalar $\phi$
and a Dirac fermion $\chi$. As well as the $U(N)$ gauge symmetry,
the Lagrangian also enjoys a $SU(N)$ flavor symmetry. Under these
two groups, the scalar multiplets transform as $(\bf{N,\bar N})$.
Thus, $\phi$ and $\chi$ fields can be seen as $N\!\times\! N$
matrices $\phi=\phi^i_{\;a}$ and $\chi=\chi^i_{\;a}$, where the
indices $i,j,...=1,2,...,N$ refer to the gauge group and
$a,b,...=1,2,...,N$ to the flavor group.

We represent the gauge fields in terms of matrices in the
fundamental representation of $U(N)$, that is, $A_\mu = A_\mu^r
T^r$, where $T^r$ ($r,s,...=1,2,..., N^2$) are the generators of
the ${\bf N}$ representation. Following the notation of
\cite{Nishino:1991sr}, we take the space-time metric
$\eta_{\mu\nu}={\rm diag}(+1,-1,-1)$, $\epsilon^{012}=+1$ and use
anti-Hermitian generators $T^r$ satisfying
\be
   [(T^r)_{\; j}^i]^*=-(T^r)_{\; i}^j, \;\;\;\;\;\;\;\;\;\;\;\;\;
   [T^r,T^s]=f^{rst}T^t, \;\;\;\;\;\;\;\;\;\;\;\;\; (T^rT^s)_{\;
   i}^i=-\frac 1 2 \delta^{rs}.
\ee

Once the auxiliary fields are eliminated by means of the field
equations, the Lagrangian for a ${\cal N}=2$ Chern-Simons-Higgs
system without holomorphic superpotential takes the form
\cite{Nishino:1991sr},\cite{Gates:1991qn}
\bea
    {\cal L}&=&  \frac{\kappa}{2}\epsilon^{\mu\nu\rho}\left(F^r_{\mu\nu}A^r_\rho
    -\frac{e}{3}f^{rst}A^r_\mu A^s_\nu A^t_\rho\right)
    + {\cal D}_\mu \bar\phi^a{\cal D}^\mu \phi_a
    + \frac i 2 \bar\chi^a\not\!\! {\cal D} \chi_a\nonumber\\
    && -\frac{e^2}{4\kappa}(\bar\phi^aT^r\phi_a-i\xi^r)(\bar\chi^bT^r\chi_b)-
    \frac {e^2}{2\kappa}(\bar\chi^a T^r \phi_a)(\bar\phi^b T^r
    \chi_b)\nonumber\\
    && - \frac  {e^4}{8\kappa^2}(\bar\phi^a\{T^r, T^s\}\phi_a)(\bar\phi^bT^r\phi_b-i\xi^r)
    (\bar\phi^cT^s\phi_c-i\xi^s)
    \label{lag}
\eea
where
\bea
    \bar\phi^a_{\;i}\equiv(\phi^i_{\;a})^*,\;\;\;\;\;\;\;
    \bar\chi^a_{\;i}\equiv(\chi^i_{\;a})^*,\;\;\;\;\;\;\;
    F^r_{\mu\nu}=\partial_\mu A_\nu^r-\partial_\nu A_\mu^r
    +ef^{rst}A_\mu^s A_\nu^t
\eea
and
\be
    {\cal D}_\mu \phi^i_{\;a} \equiv \partial_\mu \phi^i_{\;a}
    + eA_\mu^r (T^r)_{\;j}^i \phi^j_{\;a},\;\;\;\;\;\;\;
    {\cal D}_\mu \bar\phi^a_{\;i} \equiv \partial_\mu \bar\phi^a_{\;i}
    - eA_\mu^r (T^r)_{\;i}^j \bar\phi^a_{\;j},\;\;\;\;\;\;\;
    \not\!\!{\cal D} \chi^i_{\;a} \equiv {\not\!\partial} \chi^i_{\;a}
    + e{\not\!\!A}^r (T^r)_{\;j}^i \chi^j_{\;a}.
\ee
For simplicity we do not write  gauge group indices which
are summed, e.g.,
\be
   (\bar\phi^aT^r\phi_a)\equiv(T^r)_{\;j}^i\bar\phi^a_{\;i}\phi^j_{\;a},\;\;\;\;\;
   (\bar\chi^aT^r\chi_a)\equiv(T^r)_{\;j}^i\bar\chi^a_{\;i}\chi^j_{\;a},\;\;\;\;\;{\rm etc.}
\ee
Besides, all the barred spinors will denote
complex conjugates of the unbarred ones (see the Appendix for our
spinor conventions).

The Fayet-Iliopoulos parameters $\xi^r$ are nonvanishing only for
the $U(1)$ factor group in the total gauge group (we take
$\xi^{r=1}\equiv\sqrt{N/2}\,\xi$). As usual for
non-Abelian C-S theories, a consistent (gauge invariant) quantization
requires the
coefficient $\kappa$ to be
\cite{Deser:1982vy}:
\be
   \kappa= \frac{me^2}{8\pi} \;\;\;\;(m=\pm1,\pm2,...).
\ee
Note that the renormalizable sixth-order potential for the scalar
field is not related to the superpotential (which in our model is
absent) and it is uniquely determined by the ${\cal N}=2$
supersymmetry of the model.

The equations of motion obtained from the variation of the action
are
\bea
    {\cal D}_\mu {\cal D}^\mu \phi_a &=&
    -\frac{e^2}{4\kappa}T^r\phi_a(\bar\chi^bT^r\chi_b)+
    \frac {e^2}{2\kappa}T^r \chi_a(\bar\chi^b T^r \phi_b)
    - \frac  {e^4}{8\kappa^2}\{T^r, T^s\}\phi_a(\bar\phi^bT^r\phi_b-i\xi^r)
    (\bar\phi^cT^s\phi_c-i\xi^s)\nonumber\\
    && - \frac  {e^4}{4\kappa^2}T^r\phi_a(\bar\phi^b\{T^r, T^s\}\phi_b)
    (\bar\phi^cT^s\phi_c-i\xi^s),\\
    i \not\!\! {\cal D} \chi_a &=&
    \frac{e^2}{2\kappa}T^r\chi_a(\bar\phi^bT^r\phi_b-i\xi^r)+
    \frac {e^2}{\kappa}T^r \phi_a(\bar\phi^b T^r
    \chi_b),\\
    \kappa \epsilon^{\mu\nu\rho}F^r_{\nu\rho} &=&  e\vphantom{ \frac{1}{2\kappa}}J^{r\mu},
    \label{equ}
\eea
where the conserved matter current $J^{r\mu}=(\rho^r,{\bf J}^r)$
is given by
\be
   J^{r\mu}= {\cal D}^\mu\bar\phi^aT^r\phi_{a}-\bar\phi^a T^r{\cal
   D}^\mu\phi_{a} + \frac {i}{2}\bar\chi^a\gamma^\mu T^r\chi_{a}.
\ee

The $0$-component of eq.(\ref{equ}),
\be
    2\kappa B^r = e\rho^r
    \label{gauss}
\ee
is just the Chern-Simons version of the Gauss law and implies that
any object carrying magnetic flux must also carry electric charge.

The ${\cal N}=2$ supersymmetry transformation for the Lagrangian
(\ref{lag}) are
\bea
    \delta A_\mu^r&=& \frac{ie}{2\kappa}\left(\bar \epsilon \gamma_\mu
    \bar \phi^aT^r\chi_a + \epsilon \gamma_\mu
    \bar \chi^aT^r\phi_a\right),\nonumber\\
    \delta \phi_a &=& \bar \epsilon\chi_a,\;\;\;\;\;\;\;\;\;\;\;\;\;\;\;\;
    \delta \bar\phi^a \,=\,  \epsilon\bar\chi^a,\nonumber\\
    \delta \chi_a&=& -2i\gamma^\mu\epsilon{\cal D}_\mu\phi_a +
    \frac {e^2} {\kappa} \epsilon T^r \phi_a(\bar\phi^bT^r\phi_b-i\xi^r),\nonumber\\
    \delta \bar \chi^a&=& -2i\gamma^\mu\bar\epsilon{\cal D}_\mu\bar\phi^a
    +
    \frac {e^2} {\kappa} \bar\epsilon \bar\phi^aT^r
    (\bar\phi^bT^r\phi_b-i\xi^r).
    \label{trans}
\eea

A supersymmetric vacua of the theory exists whenever the minima of
the
potential can be set to zero. This happens for constant scalar
field configurations satisfying
\be
    0=\left(\bar\phi^a_{\; k}(T^r)^k_{\;l}\phi_{\;a}^l-i\xi^r\right)(T^r)^i_{\;j}
    \phi_{\;b}^j=-\frac 1 2 \left(\phi^i_{\;a}\bar\phi^a_{\;
    j}-\xi\delta^i_j\right)\phi_{\;b}^j,
    \label{condi2}
\ee
To get the last term it is helpful the relation satisfied by
the matrices of the fundamental representation of $U(N)$,
\be
   (T^r)_{\;j}^i(T^r)_{\;l}^k=-\frac{1}{2}\delta_{l}^i\delta_{j}^k.
\ee
From eq.(\ref{condi2}) one can see that there are two types of
vacua. The (gauge) symmetric phase, where
\be
   \phi^i_{\;a}=0
\ee
and the asymmetric phase, where

\be
   \phi^i_{\;a}\bar\phi^a_{\;j} =\xi \delta^i_j.
   \label{condi}
\ee
It is clear that there is only one vacua in the asymmetric phase
which, up to gauge rotations, takes the form
\be
   \phi^i_{\;a}=\sqrt{\xi} \delta^i_a.
   \label{vac}
\ee
The vacuum field (\ref{vac}) has the pattern of symmetry breaking
\cite{Hanany:2003hp}-\cite{Auzzi:2003fs}
\be
    U(N)_{\rm color}\times SU(N)_{\rm flavor} \longrightarrow SU(N)_{\rm c+f},
\ee
where the surviving unbroken group $SU(N)_{\rm c+f}$ is a simultaneous
gauge and flavor rotation. Due to this, the theory is said to lie
in the colour-flavor locked phase.

\section{Non-Abelian Vortices}

\subsection{Bogomol'nyi Bound and Self-Duality Equations}

The Hamiltonian for the bosonic sector of the theory is
\be
   {\cal H} = \int\,d^2x \left( {\cal D}_0 \bar\phi^a {\cal D}_0 \phi_a
   + \overrightarrow{{\cal D}} \bar\phi^a\cdot \overrightarrow{{\cal D}} \phi_a
   + \frac  {e^4}{8\kappa^2}(\bar\phi^a\{T^r, T^s\}\phi_a)(\bar\phi^bT^r\phi_b-i\xi^r)
    (\bar\phi^cT^s\phi_c-i\xi^s)\right)
\ee

One can find the first order vortex equations by the usual
Bogomol'nyi trick
\cite{Bogomolny:1975de}. That is, using the relation
\be
   \overrightarrow{{\cal D}} \bar\phi^a \cdot\overrightarrow{{\cal D}}
   \phi_a = ({\cal D}_1\mp i{\cal D}_2)\bar\phi^a
   ({\cal D}_1\pm i{\cal D}_2)\phi_a \pm i eB^r \bar\phi^a T^r
   \phi_a \mp i {\rm tr}\,(\partial_1\, J^{r}_2 -\partial_2\, J^{r}_1) T^r
   \label{complet}
\ee
and the Gauss law eq.(\ref{gauss}), the Hamiltonian can be written
as
\bea
   {\cal H} &=& \int\,d^2x \left\{ \left({\cal D}_0 \bar\phi^a \mp
   \frac {ie^2}{2\kappa}
   (\bar\phi^bT^r\phi_b-i\xi^r)\bar\phi^a T^r\right)\left({\cal D}_0 \phi_a \pm \frac {ie^2}{2\kappa}
   (\bar\phi^cT^s\phi_c-i\xi^s)T^s\phi_a \right)\right.\nonumber\\
   && \left. \vphantom{\frac {ie^2}{2\kappa}} + ({\cal D}_1\mp i{\cal D}_2)\bar\phi^a
   ({\cal D}_1\pm i{\cal D}_2)\phi_a\right\} \pm  e \xi \Phi
   \label{cota}
\eea
The integral of the last term in the r.h.s of eq.(\ref{complet})
does not contribute to the energy since it can be written as  a surface
term vanishing at spatial infinity for any finite-energy solution.
On the other hand, $\Phi$ is the topological charge given by
\be
   \Phi=-i\int d^2x\,{\rm tr} B
   \label{mag}
\ee
which coincides
with the $U(1)$ magnetic flux.

The energy is then bounded according to
\be
   {\cal H}\geq e\xi |\Phi|
   \label{cota2}
\ee
It is clear from eq.(\ref{cota}) that the bound is saturated by
configurations satisfying the Gauss law and the self-duality
(Bogomol'nyi) equations:

\bea
   &&{\cal D}_0 \phi_a - \frac {ie^2}{2\kappa}
   \varepsilon(\bar\phi^bT^r\phi_b-i\xi^r)T^r\phi_a=0\nonumber\\
   &&({\cal D}_1 -  i\varepsilon{\cal D}_2)\phi_a =0
   \label{bogo}
\eea
with $\varepsilon=\pm$ being the sign of the topological charge.
Since static configurations that are stationary points of the
energy are also stationary points of the action, the
Euler-Lagrange equations of the theory will be satisfied by such static
configurations obeying the Gauss law and the self-duality
equations.

As it is well known, the existence of a Bogomol'nyi bound for the
energy is strongly related to the ${\cal N}=2$ supersymmetry of
the theory \cite{witten}. Indeed, to require that the Bogomol'nyi bound
(\ref{cota2}) be saturated  is equivalent to look for
configurations invariant under half of the supersymmetry
transformations (\ref{trans}). In order to verify this, we define
the parameters
\be
   \epsilon^\pm = \frac 1 2 \left( 1 \pm \varepsilon\gamma^0\right) \epsilon.
\ee
The supersymmetric variation of the fermion field
$\chi$ in a bosonic background satisfying
eq.(\ref{bogo})
results
\be
   \delta \chi_a = -2i(\gamma^\mu{\cal D}_\mu\phi_a
    +\varepsilon{\cal D}_0\phi_a)\epsilon^+
\ee
and then the $\chi$ field remains invariant under any supersymmetry transformation with
$\epsilon^+=0$. Concerning the supersymmetry variation of the bosonic fields, they
are automatically zero since we start with a purely bosonic
classical configuration. Thus, self dual solutions are 1/2 BPS,
with $\epsilon^+$ the parameters of the broken supersymmetry and
$\epsilon^-$ the parameters of the unbroken supersymmetry.

\subsection{Non-Abelian Multisoliton Solution}

In order to find non-Abelian vortices in this model, let us
consider rotationally symmetric configurations through the ansatz \cite{Auzzi:2003fs}:
\bea
     \phi &=& \left(
          \begin{array}{ccccc}
          \varphi(r) & 0 & \cdots & 0 & 0 \\
           0& \varphi(r) & \cdots & 0& 0 \\
          \vdots & \vdots & \ddots & \vdots\\
           0 & 0& \cdots & \varphi(r)& 0 \\
           0& 0 & \cdots & 0& e^{in\theta}\tilde\varphi(r) \\
          \end{array}
          \right),\nonumber\\
     \nonumber\\
     A_\alpha^{\rm SU(N)} &=& \left(
          \begin{array}{ccccc}
          \;\;\,1\;\; & 0& \cdots & 0& 0 \\
           0& \;\;\,1 \;\; & \cdots & 0 & 0\\
          \vdots & \vdots & \ddots &\vdots & \vdots\\
           0 & 0& \cdots & \;\;\,1 \;\;& 0 \\
           0& 0 & \cdots & 0& -(N-1)\\
          \end{array}
          \right)\frac{i}{Ne}(\partial_\alpha \theta) [n-f_N(r)], \nonumber\\
     \nonumber\\
     A_0^{\rm SU(N)} &=& \left(
          \begin{array}{ccccc}
          \;\;\,1\;\; & 0& \cdots & 0& 0 \\
           0& \;\;\,1 \;\; & \cdots & 0 & 0\\
          \vdots & \vdots & \ddots &\vdots & \vdots\\
           0 & 0& \cdots & \;\;\,1 \;\;& 0 \\
           0& 0 & \cdots & 0& -(N-1)\\
          \end{array}
          \right)ig_N(r),\nonumber\\
     \nonumber\\
     A_\alpha^{\rm U(1)} &=& \sqrt{\frac{2}{N}}\frac 1 e (\partial_\alpha \theta) [-n+ f(r)], \nonumber\\
     \nonumber\\
     A_0^{\rm U(1)} &=& \sqrt{2N} g(r),
     \label{rota}
\eea
where $\alpha=1,2$ labels the space coordinates and $(r,\theta)$
are the polar coordinates in this space.

Inserting this ansatz in the Gauss law (\ref{gauss}) and the
self-duality equations (\ref{bogo}), we arrive to the first-order
differential equations satisfied by the profile functions
\bea
    &&r\partial_r \varphi+\frac {\varepsilon} {N} (f-f_N)\varphi
    =0\nonumber\\
    &&r\partial_r \tilde\varphi+\frac {\varepsilon} {N} (f-(1-N)f_N)\tilde\varphi
    =0\nonumber\\
    &&\frac 1 r \partial_r (f-f_N) - \varepsilon\frac{N
    e^4}{8\kappa^2}\varphi^2(\varphi^2-\xi)=0\nonumber\\
    &&\frac 1 r \partial_r (f-(1-N)f_N) - \varepsilon\frac{N
    e^4}{8\kappa^2}\tilde\varphi^2(\tilde\varphi^2-\xi)=0\nonumber\\
    && g_N = \frac {\varepsilon e}{4N\kappa}(\tilde\varphi^2-\varphi^2)\nonumber\\
    && g = -\frac {\varepsilon e}{4N\kappa}(\tilde\varphi^2-(1-N)\varphi^2-N\xi)
    \label{prof eq}
\eea

The boundary conditions at the origin follows from the requirement
that the fields be nonsingular. This implies that
\be
   n\tilde\varphi(0)=0,\quad f_N(0)=n,\quad f(0)=n.
\ee
At spatial infinity, finiteness of the energy implies that
\be
   \varphi(\infty)=\tilde\varphi(\infty)=0\;{\rm or}\;\sqrt
   \xi,\quad \varphi(\infty)f (\infty)=\varphi(\infty)f_N (\infty)=0
\ee
In searching for solutions of eqs.(\ref{prof eq}), we shall consider
only those with $\Phi>0$. Solutions with negative magnetic flux
are related to these ones by the transformation
\be
   \varphi\rightarrow\varphi,\quad\tilde\varphi\rightarrow\tilde\varphi,\quad
   f\rightarrow-f,\quad f_N\rightarrow-f_N,\quad g\rightarrow-g,\quad
   g_N\rightarrow-g_N.
\ee

Note that, with this ansatz, the magnetic flux (\ref{mag}) and the
$U(1)$ electric charge take the form
\be
   \Phi=  \sqrt{\frac{N}{8}}\frac{e}{\kappa}\,Q^{\rm U(1)}
   =\frac{2\pi}{e}
   (f(0)-f(\infty))=\frac{2\pi}{e}(n+\alpha)
   \label{flux}
\ee
where we have written $f(\infty)\equiv-\alpha$, being $\alpha=0$
when $\varphi(\infty)\ne 0$
or an
undetermined constant when $\varphi(\infty)=0$.

In order to get a better knowledge about these solutions,
it will be convenient to define new profile functions $h(r)$ and
$\tilde h(r)$ as:
\be
   h(r)=\frac 1 N (f(r)-f_N(r)),\quad
   \tilde h(r)= \frac 1 N (f(r)-(1-N)f_N(r)).
\ee
Now, in terms of these functions, eqs.(\ref{prof eq}) reduce to two
sets of equations, each one being identical to that appearing in
the case of the Abelian self-dual Chern-Simons soliton
\cite{hong},\cite{Jackiw:1990pr}. That is, a first set of equations for the
functions $(\varphi, h)$
\be
   r\partial_r \varphi + h\varphi = 0,\quad\quad
   \frac 1 r \partial_r h - \frac{e^4}{8\kappa^2}
   \varphi^2(\varphi^2-\xi)=0
   \label{1}
\ee
with the boundary conditions
\be
   h(0)=0,\quad \varphi(\infty)=0\;{\rm or}\;\sqrt
   \xi,\quad \varphi(\infty)f (\infty)=0,
\ee
and an identical set of equations for the functions
$(\tilde\varphi,\tilde h)$
\be
   r\partial_r \tilde\varphi +\tilde h\tilde\varphi = 0,\quad\quad
   \frac 1 r \partial_r \tilde h - \frac{e^4}{8\kappa^2}
   \tilde\varphi^2(\tilde\varphi^2-\xi)=0
   \label{2}
\ee
but with the less restrictive boundary conditions
\be
   n\tilde\varphi(0)=0,\quad \tilde h(0)= n,\quad
   \tilde\varphi(\infty)=0\;{\rm or}\;\sqrt
   \xi,\quad \tilde\varphi(\infty)\tilde h(\infty)=0.
\ee

As shown in \cite{hong}, these equations admit vortex like
solutions for which $\varphi\rightarrow\sqrt \xi$ and
$\tilde\varphi\rightarrow\sqrt \xi$ at large distances. They are
topologically nontrivial configurations having, as one can see
from eq.(\ref{flux}), quantized magnetic flux $\Phi=2\pi n/e$.
Hence, they cannot be continuously deformed to the vacuum solution
for topological reasons. There also exist nontopological soliton
solutions for which the Higgs field $\phi$ approaches the
symmetric minimum at large distances \cite{Jackiw:1990pr}. Their
flux is not quantized, but rather it is expressed in terms of the
arbitrary parameter $\alpha$ describing the solution. The rest of
the paper we will be focus in the study of configurations of the
former type, that is, solutions in the topological sector. In this
case, eqs.(\ref{1}),(\ref{2}) admit vortex-like solutions for
every non vanishing entire $n$. As we will see, these solutions
give truly non-Abelian generalization of the Abelian vortices
first discussed in \cite{hong}.

An important simplification of the ansatz (\ref{rota}) arises as
a result of the fact that the vacuum solution
\be
   \varphi(r)\equiv\sqrt \xi,\quad h(r)\equiv0,
\ee
is the only solution of eqs.(\ref{1})\cite{Jackiw:1990pr}. Therefore, the Higgs field
and the gauge fields in eq.(\ref{rota}) take the simple form
\bea
    \phi &=& {\rm diag}(\sqrt \xi, \sqrt \xi, ... , \sqrt \xi, e^{in\theta}
    \tilde\varphi(r))\vphantom{\frac {ie}{4\kappa}}\nonumber\\
    A_\alpha &=& {\rm diag}(0, 0, ... , 0, 1)\frac{i}{e}\partial_\alpha \theta(f(r)-n),\quad\alpha=1,2
    \nonumber\\
    A_0 &=& {\rm diag}(0, 0, ... , 0, 1)\frac {ie}{4\kappa}(\xi-\tilde\varphi^2)
    \label{simple}
\eea
where $A_\mu = A_\mu^{\rm SU(N)} + \frac {i}{\sqrt 2N} A_\mu^{\rm
U(1)}I$, $\mu=0,1,2$. Besides, the profile functions
$(\tilde\varphi(r),f(r))$ satisfy the following first-order equations and boundary conditions
\bea
   &&r\partial_r \tilde\varphi + f\tilde\varphi = 0,\quad\quad
   \frac 1 r \partial_r f - \frac{e^4}{8\kappa^2}
   \tilde\varphi^2(\tilde\varphi^2-\xi)=0,\nonumber\\
   && n\tilde\varphi(0)=0,\quad f(0)= n,\quad
   \tilde\varphi(\infty)=\sqrt
   \xi,\quad f(\infty)=0
\eea
and then, coincide with the profiles of an Abelian self-dual
Chern-Simons vortex with $n$ units of topological charge
\cite{hong}.

Let us now discuss some facts about the vortex moduli space. While
the vacuum in the asymmetric phase is $SU(N)_{\rm c+f}$ symmetric,
the solution given by eq.(\ref{rota}) breaks this symmetry down to
$U(1)\times SU(N-1)$. This means that there exist a set of
solutions with the same topological charge parameterized by the coset
\cite{Hanany:2003hp},\cite{Auzzi:2003fs}
\be
   \frac{SU(N)_{\rm c+f}}{SU(N-1)\times U(1)} \cong {\bf CP}^{N-1}
\ee

Thus, in the case of a single unit charge vortex the moduli space
decomposes as
\be
   {\cal M}\cong {\bf C}\times{\bf CP}^{N-1}
\ee
where ${\bf C}$ parameterizes the center of mass of the vortex
configuration. The presence of these extra orientational
collective coordinates makes the vortices genuinely non-Abelian.

Concerning charge-$n$ vortices, one expects from the BPS nature of
these configurations that there exist solutions corresponding to
$|n|$ well-separated unit charge vortices. Then, as usually
happens with solitons satisfying a set of Bogomol'nyi-type
equations, the charge-$n$ vortex moduli space should involve as
many independent parameters as those necessary to determine the
state of $|n|$ independent charge-$1$ vortices. We have seen above
that each charge-$1$ vortex is characterized by its position on the
plane together with $(N-1)$ complex parameters describing the
orientation of the vortex in the $SU(N)_{\rm c+f}$ group.
Therefore, the dimension of the charge-$n$ vortex moduli space
will be $2|n|N$.

One can  make explicit the non-Abelian nature of the solution
(\ref{rota}) by applying the color-flavor rotation preserving the
asymmetric vacuum. To this end, it is convenient first to pass to
the singular gauge where the scalar fields have no winding at
infinite, while the vortex flux comes from the vicinity of the
origin. Then, the Higgs and the gauge fields can be written as
\bea
     \phi &=& {\cal U} \left(
          \begin{array}{ccccc}
          \sqrt \xi & 0 & \cdots & 0 & 0 \\
           0& \sqrt \xi & \cdots & 0& 0 \\
          \vdots & \vdots & \ddots & \vdots& \vdots \\
           0 & 0& \cdots & \sqrt \xi\\
           0& 0 & \cdots & 0& \tilde\varphi(r) \\
          \end{array}
          \right){\cal U}^{-1},\nonumber\\
     \nonumber\\
     A_\alpha^{\rm SU(N)} &=& -{\cal U}\left(
          \begin{array}{ccccc}
          \;\;\,1\;\; & 0& \cdots & 0& 0 \\
           0& \;\;\,1 \;\; & \cdots & 0 & 0\\
          \vdots & \vdots & \ddots &\vdots & \vdots\\
           0 & 0& \cdots & \;\;\,1 \;\;& 0 \\
           0& 0 & \cdots & 0& -(N-1)\\
          \end{array}
          \right){\cal U}^{-1}\frac{i}{Ne}(\partial_\alpha \theta) f(r), \nonumber\\
     \nonumber\\
     A_0^{\rm SU(N)} &=& {\cal U}\left(
          \begin{array}{ccccc}
          \;\;\,1\;\; & 0& \cdots & 0& 0 \\
           0& \;\;\,1 \;\; & \cdots & 0 & 0\\
          \vdots & \vdots & \ddots &\vdots & \vdots\\
           0 & 0& \cdots & \;\;\,1 \;\;& 0 \\
           0& 0 & \cdots & 0& -(N-1)\\
          \end{array}
          \right){\cal U}^{-1}\frac {i\varepsilon e}{4N\kappa}(\tilde\varphi^2-\xi),\nonumber\\
     \nonumber\\
     A_\alpha^{\rm U(1)} &=& \sqrt{\frac{2}{N}}\frac 1 e (\partial_\alpha \theta) f(r), \nonumber\\
     \nonumber\\
     A_0^{\rm U(1)} &=& -\frac {\varepsilon e}{\sqrt{8N}\kappa}(\tilde\varphi^2-\xi),
     \label{rota2}
\eea
where ${\cal U}\in SU(N)$ parameterizes the orientational
collective coordinates associated with the flux rotation in
$SU(N)$.

\section{Effective vortex world-line theory}

In this section we will derive an effective low-energy theory for
the orientational collective coordinates on the vortex world-line.
Following \cite{Gorsky:2004ad}, we parameterize the matrices
entering in eq.(\ref{rota2}) as follows:
\be
   \frac{1}{N}\left\{{\cal U}\left(
          \begin{array}{ccccc}
          \;\;\,1\;\; & 0& \cdots & 0& 0 \\
           0& \;\;\,1 \;\; & \cdots & 0 & 0\\
          \vdots & \vdots & \ddots &\vdots & \vdots\\
           0 & 0& \cdots & \;\;\,1 \;\;& 0 \\
           0& 0 & \cdots & 0& -(N-1)\\
          \end{array}
          \right){\cal U}^{-1}\right\}^i_{\, j}=-n^i n_j^* + \frac 1 N
          \delta^i_j,
\ee
where $n^i$ is a complex vector in the fundamental representation
of $SU(N)$, and
\be
   n^*_i n^i = 1\quad\quad i=1,...,N
   \label{const}
\ee
Note that this gives the correct number of degrees of freedom for
the charge-$1$ vortex case, namely, $2(N-1)$.

With this parametrization the vortex solution (\ref{rota2}) takes
the form
\bea
    \phi &=& \frac 1 N
    [\tilde\varphi+(N-1)\sqrt\xi]+(\tilde\varphi-\sqrt\xi)\left(n\cdot n^* - \frac 1
    N\right)\nonumber\\
    A_\alpha^{SU(N)}&=&\frac i e \left(n\cdot n^* - \frac 1
    N\right)\partial_\alpha\theta f(r)\nonumber\\
    A_\alpha^{U(1)}&=&\sqrt{\frac 2 N}\frac 1 e \partial_\alpha\theta f(r)\nonumber\\
    A_0^{SU(N)}&=&-\frac {i\varepsilon e}{4\kappa}(\tilde\varphi^2-\xi)\left(n\cdot n^* - \frac 1
    N\right)\nonumber\\
    A_0^{U(1)}&=& \frac {\varepsilon e}{\sqrt{8 N}\kappa}(\tilde\varphi^2-\xi)
\eea
where $\alpha=1,2$. For simplicity we have suppressed all $SU(N)$
indices.

To obtain the low-energy effective theory, we assume that
the moduli
$n^i$ are slowly-varying functions of the $x^0$-coordinate,
promoting the $2(N-1)$ collective coordinates to dynamical fields
on the vortex world-line. Plugging the resulting configuration
into the Lagrangian (\ref{lag}) and performing the integral over
the $(x^1,x^2)$ plane we finish with a one-dimensional sigma-model
for the bosonic $n^i$ coordinates. However, before doing this we
have to modify our solution. The point is that our solution was
obtained through a color-flavor rotation, which now gets a
dependence on the $x^0$-coordinate. Therefore, the 0-component of
the gauge potential has to be modified. Following an ansatz
similar to that used in \cite{Gorsky:2004ad}, we propose
\bea
    A_0^{SU(N)}&=&-\frac {i\varepsilon \,e}{4\kappa}(\tilde\varphi^2-\xi)\left(n\cdot n^* - \frac 1
    N\right)-\frac 1 e[\partial_0n \cdot n^* - n \cdot \partial_0n^*
    -2 n\cdot n^*(n^* \partial_0n)]\rho(r)\nonumber\\
    A_0^{U(1)}&=&\frac {\varepsilon\, e}{\sqrt{8 N}\kappa}(\tilde\varphi^2-\xi)
\eea
where we have introduced a new profile function $\rho(r)$. It will be
determined by its own equation of motion through a minimization
procedure.

The kinetic term for $n^i$ comes from the gauge and scalar field
kinetic terms in eq.(\ref{lag}), while, due to $n^i$ parameterize
the vortex zero modes, no potential term is expected to be present
in this sigma-model.

Using the latter ansatz for the gauge potential we can calculate
the $U(N)$ gauge field strength, which takes the form
\bea
   F_{0 \alpha} &=& -\frac i e(\partial_0 n \cdot n^* +  n \cdot \partial_0n^*)
   \varepsilon_{\alpha\beta} \frac{x^\beta}{r^2} f(r) (1-\rho(r))
   + \frac {i\varepsilon e}{4\kappa} \frac{x_\alpha}{r}\partial_r
   \tilde\varphi^2(r) \,n\cdot n^*  \nonumber\\
   &&+\frac 1 e[\partial_0n \cdot n^* - n \cdot \partial_0n^*
    -2 n\cdot n^*(n^* \partial_0 n)]\frac{x_\alpha}{r}\partial_r\rho(r)
\eea
and the other components zero.

Therefore, the Chern-Simons term in the Lagrangian (\ref{lag})
is such that
\bea
    \epsilon^{\mu\nu\rho} (F^r_{\mu\nu}A^r_\rho
    -\frac{e}{3}f^{rst}A^r_\mu A^s_\nu A^t_\rho)=
    -\frac {\varepsilon} {\kappa r} f \partial_r\tilde\varphi^2
    \label{new}
\eea
Eq.(\ref{new}) implies that when the Chern-Simons governs the gauge field dynamics
 the only
$x^0$-dependent terms contributing to the effective Lagrangian
will come from the scalar field kinetic term. After some algebra,
one arrives at
\be
   S_{eff} = \frac {2\pi}{e^2}I_0 \int  dx^0 \,[\partial_0n^*\partial_0n+(n^*\partial_0n)^2]
   \label{lageff}
\ee
where the constant $I_0$ is given by the following integral
defined in terms of the adimensional quantities $\hat r= e^2 r$,
$\hat \varphi=e^{-1}\tilde\varphi$ and $\hat \xi=e^{-2}\xi$,
\be
   I_0 = \int_0^\infty d\hat r\,\hat r \left [(\hat \varphi^2 +
   \hat\xi)\rho^2 + 2 (1-\rho) (\hat \varphi -
   \sqrt{\hat\xi})^2\right].
\ee

Minimization of the effective action gives the equation of motion
of $\rho$. Note that, contrary to what happens in Yang-Mills
theories, in this case $\rho$ results a non-dynamical field. Then,
it can be easily put in terms of $\tilde\varphi$, taking the form
\be
    \rho = \frac{(\tilde \varphi - \sqrt\xi)^2}
    {\tilde \varphi^2 + \xi}.
\ee

Due to action (\ref{lageff}) is invariant under
the $U(1)$ gauge transformation
\be
   n^i \rightarrow e^{i\vartheta(t)} n^i,\quad
   n^*_i \rightarrow e^{-i\vartheta(t)} n^*_i
\ee
and fields $n^i$ are subject to the constraint (\ref{const}), the effective
world-line theory can be recognized as the one
dimensional ${\bf CP}^{N-1}$ theory, as was already anticipated by
using symmetry arguments.

\section{Quantization of the one-dimensional ${\cal N}=2$ ${\bf CP}^{1}$ model}

We shall here proceed to the analysis of the model defined by
eq.(\ref{lageff}) in the case of $N=2$ both at the classical and
quantum level as a way of describing the low energy properties of
the non-Abelian $U(2)$ Chern-Simons vortices. As we have seen in
the latter section, the dynamics of the two collective coordinates
$n^1(t),n^2(t)$ is given by a one-dimensional ${\bf CP}^{1}$
theory whose Lagrangian takes the form
\be
   {\cal L}_B = \frac {2\pi }{e^2}I_0\,[\partial_tn^*\partial_tn+(n^*\partial_tn)^2]
   \label{lageff1}
\ee
and with the coordinates $n^1,n^2$ satisfying the constraint
\be
    n^*n=n^*_1n^1+n^*_2n^2=1
    \label{const1}
\ee
In order to quantize this theory, it will be convenient to solve
first the constraint eq.(\ref{const1}) through the stereographic
projection and deal with the holomorphic representation of the
Lagrangian (\ref{lageff1}). The holomorphic representation is
defined in terms of one complex field $z(t)$ and its complex
conjugated $\bar z(t)$, which are functions from time onto the
Kahler manifold ${\bf CP}^{1}$. In this way, the theory describing
the dynamics of the bosonic orientational moduli results an
unconstrained one dimensional ${\bf CP}^{1}$ sigma model. Thus,
the effective Lagrangian can be written as
\bea
   {\cal L}_B= g_{z \bar z} \dot z \dot {\bar z}
   \label{lageff2}
\eea
where $\dot z=\partial_t z$ and $g_{z \bar z}$ is the ${\bf
CP}^{1}$ metric
\be
   g_{z \bar z}=\frac {r_0^2}{(1+z\bar z)^2}.
\ee
We have renamed the constant coupling as $r_0^2$, so that $r_0 \propto
e^{-1}$. In the case of ${\bf CP}^{1}$ target space, this implies
that the scalar curvature $R$ results $R \propto e^2$. Note that due
to the Kahler condition of this manifold, $g_{z \bar z}$ can be
derived from a Kahler potential, that is, $g_{z \bar
z}=\partial_z\bar\partial_{\bar z}{\cal K}$ with ${\cal K}=r_0^2
\log(1+z\bar z)$.

Moreover, since vortices remain invariant under the action of one
half of the supersymmetries, that is, under one complex
supercharge, the world-line theory must have ${\cal N}=2$
supersymmetries. Thus, we can use the unbroken supersymmetry to
reconstruct the fermionic sector of the theory. This implies
that the quantum mechanics of the orientational moduli in the
low-energy regime is given by a one-dimensional supersymmetric ${\cal N}=2$
${\bf CP}^{1}$ sigma model. The Lagrangian for this
theory is
\bea
   {\cal L}= g_{z \bar z} \dot z \dot {\bar z}+\frac i 4  g_{z \bar z} \left(\bar \psi {\cal
   D}_t\psi- \bar{\cal D}_t\bar \psi \psi \right)
   \label{lagri}
\eea
where $\psi(t)$ is a complex Grassmann variable and $\bar \psi(t)$
its complex conjugated. Besides, ${\cal D}_t$ and $\bar{\cal D}_t$
are the covariant derivatives
\bea
   {\cal D}_t\psi = \dot \psi + \Gamma\dot z\psi,\quad \quad\bar{\cal
   D}_t\bar\psi = \dot {\bar\psi} + \bar\Gamma\dot{\bar z}\bar\psi,
\eea
with $\Gamma=g^{\bar z z}\partial_z g_{z\bar z}$ and
$\bar\Gamma=g^{\bar z z}\bar\partial_{\bar z} g_{z\bar z}$ being
the Christoffel symbols.
Lagrangian (\ref{lagri}) results invariant under the ${\cal N}=2$
supersymmetry transformation
\bea
    \delta z \!&=&\! \epsilon \psi, \;\;\;\;\;\;\;\;\;\;\;\;\;\; \delta \psi
    = -2i \bar\epsilon \dot z, \nonumber\\
    \delta \bar z \!&=&\!  \bar\psi\bar\epsilon, \;\;\;\;\;\;\;\;\;\;\;\;\;\;
    \delta \bar\psi = 2i \epsilon \dot{\bar z}.
    \label{susy}
\eea

In order to proceed to the (canonical) quantization of the theory we
find, from (\ref{lagri}), the canonical momenta
\bea
    p &\equiv& \frac{\partial {\cal L}}{\partial \dot z}= g_{z \bar z} \dot
    {\bar z} + \frac i 4 g_{z \bar z, z} \bar \psi \psi,\quad\quad \pi \equiv
    \frac{\partial_l {\cal L}}{\partial \dot
    {\psi}}= -\frac i 4 g_{z \bar z} \bar\psi,\nonumber\\
    \bar p &\equiv& \frac{\partial {\cal L}}{\partial \dot {\bar z}}=
    g_{z \bar z} \dot z +\frac i 4 g_{z \bar z,\bar z} \psi\bar \psi,\quad\quad\bar
    \pi \equiv \frac{\partial_l {\cal L}}{\partial \dot
    {\bar\psi}}= -\frac i 4 g_{z \bar z} \psi.
    \label{momenta}
\eea
where $\partial_l$ denote the left derivative.

Concerning the Hamiltonian, it can be written as
\be
   {\cal H} \equiv \dot z \,p+ \dot{\bar z}\bar p + \dot \psi \pi +\dot
   {\bar\psi}\bar\pi- {\cal L} = \dot z g_{z \bar z} \dot {\bar z}
   \label{hami}
\ee

It follows immediately from (\ref{momenta}) that the canonical
formalism involves the fermionic second-class constraints
\be
   {\rm c}^1=\pi + \frac i 4 g_{z \bar z} \bar \psi,\quad
   {\rm c}^2=\bar\pi + \frac i 4 g_{z \bar z} \psi.
\ee
This requires use of the formalism of Dirac for constrained
systems \cite{Dirac:1950pj}. Starting from the naive Poisson brackets
\be
   \{z,p\}_{\rm PB}=1,\quad\{\bar z,\bar p\}_{\rm PB}=1,\quad \{\psi,
   \pi\}_{\rm PB}=-1,\quad \{\bar\psi, \bar\pi\}_{\rm PB}=-1,
\ee
one defines for any two field variables $f$ and $g$ the Dirac
brackets as
\be
   \{f,g\}_{\rm DB} \equiv \{f,g\}_{\rm PB} - \{f,{\rm c}^{a}\}_{\rm PB}
   {\rm C}_{ab}^{-1}\{{\rm c}^{b},g\}_{\rm PB},\quad\quad
   a,b=1,2
\ee
where
\bea
   {\rm C}^{ab} \equiv \{{\rm c}^{a},{\rm c}^{b}\}_{\rm PB}
\eea

Now, within the Dirac formalism, the basic Dirac brackets result
\bea
    &&\{z,p\}_{\rm DB}= \{\bar z,\bar p\}_{\rm DB}= 1, \quad
    \{\psi,\bar\psi\}_{\rm DB}=-2ig^{\bar z z},\\
    &&\{\psi,p\}_{\rm DB}=-\frac 1 2 \Gamma\psi= (\{\bar\psi,\bar p\}_{\rm
    DB})^*,\\
    &&\{\psi,\bar p\}_{\rm DB}=-\frac 1 2 \bar\Gamma\psi= (\{\bar\psi, p\}_{\rm
    DB})^*,
\eea
with all other trivial commutation relations omitted.

The ${\cal N}=2$ supercharges $Q$, $\bar Q$ corresponding to
transformations (\ref{susy}) take the form
\be
   Q=p\,\psi, \quad \quad \bar Q = \bar p\,\bar\psi.
   \label{char}
\ee
It is easy to see that these charges satisfy the ${\cal N}=2$
superalgebra
\be
   \{Q,\bar Q\}_{\rm DB}=-2i{\cal H}, \quad \quad\{Q, Q\}_{\rm DB}=  \{\bar Q,\bar Q\}_{\rm
   DB}=0
\ee

There is an obvious simpler set of canonical equations, which
separate the ordinary and Grassmann variables and which permit the
passage to explicit representation of the quantum mechanics. Thus,
introducing the tetrad $e$, $\bar e$ given by
\be
   e\bar e = g \Longrightarrow e=\bar e =\frac {r_0}{1+z\bar z},
\ee
we define
\be
   \lambda = e\psi, \quad \quad\bar\lambda=\bar e\bar\psi.
\ee
Therefore, the set $z,p,\lambda$ and its complex conjugated have
\be
   \{z,p\}_{\rm DB}= \{\bar z,\bar p\}_{\rm DB}= 1, \quad \quad
    \{\lambda,\bar\lambda\}_{\rm DB}=-2i
   \label{algebra}
\ee
as the only non-trivial canonical equations.

As usual, the passage from the classical theory to the quantum
theory is done through the replacement of real variables by
Hermitian operators together with the change of brackets
\be
   \{\,,\,\}_{\rm DB}\longrightarrow -i[\,,\,].
\ee

It is immediate to find a quantum mechanical representation of the
algebra eq.(\ref{algebra}). Firstly, we have for $p$ and $\bar p$
the result
\be
   p= e^{-1}(-i\partial_z)e,\quad \bar p= \bar e^{-1}(-i\bar \partial_{\bar
   z})\bar e.
   \label{bos}
\ee
Factors involving the tetrad are required to maintain the
Hermiticity relation between $p$ and $\bar p$ under the inner
product
\be
    \langle \Psi_1|\Psi_2 \rangle= \int d^2 z g_{z \bar z} \Psi_1^\dagger(z,\bar z) \Psi_2(z,\bar z)
    \label{produ}
\ee
where $\Psi(z,\bar z)$ denotes the ``invariant'' wave function
associated with the ket $|\Psi \rangle$. Secondly, we can employ a
$2\times 2$ matrix representation for $\psi,\bar\psi$, which thus
take the form
\be
   \psi=\sqrt 2\left(\begin{array}{cc}
              0 & 0\\
              1 & 0
         \end{array}\right),\;\;\;\;\;\;\;\;\;\;\;\;\;\;
   \bar\psi=\sqrt 2\left(\begin{array}{cc}
             0 & 1\\
             0 & 0
          \end{array}\right).
   \label{fer}
\ee

Concerning the ${\cal N}=2$ supercharge $Q$ and $\bar Q$, we
generalize the classical expression eq.(\ref{char}) by adopting
the following ordering for the corresponding quantum operators
\be
   Q \equiv\frac 1 2 \{p, \psi\}= \psi(p-\frac i 4 \Gamma),\quad
   \bar Q \equiv\frac 1 2 \{\bar p,\bar \psi\}= \bar\psi(\bar p+\frac i 4\bar \Gamma)
\ee

In order to ensure the supersymmetry of the system, we require
that the quantum Hamiltonian operator be defined by
\be
   {\cal H}\equiv\frac 1 2 \{Q,\bar Q\}
\ee
Therefore, using the expressions given for
eqs.(\ref{bos}),(\ref{fer}), the Hamiltonian takes the form
\be
   {\cal H}=\left(\begin{array}{cc}
              \bar AA & 0\\
              0 & A\bar A
         \end{array}\right).
\ee
with
\be
   A = i e^{-1}(\partial_z + \frac 3 4 \Gamma),\quad\quad \bar A = i \bar e^{-1}
   (\bar\partial_{\bar z} - \frac 1 4 \bar\Gamma)
   \label{A}
\ee

As it is well known, eigenfunctions of $\bar A A$ and $A \bar A$
are closely related due to supersymmetry. More explicitly, if
$\varphi_n$ is an eigenfunction of $A \bar A$ with a non-vanishing
eigenvalue $E_n$, then $\bar A \varphi_n$ is an eigenfunction of
$A \bar A$ with the same eigenvalue $E_n$. Concerning the case of
vanishing eigenvalue, from the fact that $A^\dagger=\bar A$ for
the inner product (\ref{produ}), one has that zero modes of $A\bar
A$ and $\bar A A$ correspond to states annihilated by $\bar A$ and
$A$, respectively. It is easy to see the operator $A$ as given by
eq.(\ref{A}) has no normalizable zero mode, and then neither do the
operator $\bar A A$. Based on these arguments, we can solve the
eigenvalue problem for the Hamiltonian operator ${\cal H}$ by
considering the following prescription
\be
   {\cal H}\Psi_n=E_n\Psi_n \Longleftrightarrow
   \Psi_n=\left( \begin{array}{c}
                   \bar A \varphi_n\\
                   \varphi_n
               \end{array}\right)
\ee
where $(E_n,\varphi_n)$ solve the eigenvalue problem for the
operator $A\bar A$, that is, $A\bar A\, \varphi_n=E_n\varphi_n$.

To determine the Hilbert space of eigenstates of the operator
$h=A\bar A$, let us rewrite it as \cite{Dunne:1991cs}
\be
   h= -e^{-1}\bar e^{-1}\left(\partial_z+\frac \Gamma 4\right)\left(\bar\partial_{\bar z}
   -\frac {\bar\Gamma} {4}\right)=-\frac{1}{2r_0^2\partial_z\bar\partial_{\bar
   z}\Theta}(\partial_z -\partial_z \Theta)(\bar\partial_{\bar z}
   + \bar\partial_{\bar z}\Theta)
   \label{hamilto}
\ee
with $\Theta = \kappa/2r_0^2=\log\sqrt{1 + z\bar z}$. Given an
operator of the form (\ref{hamilto}), it is natural to define the
``reduced'' wavefunctions $\hat \varphi$,
\be
   \varphi (z, \bar z)\equiv e^{-\Theta}\,
   \hat\varphi (z, \bar z)
\ee
 Thus, the ``reduced'' operator $\hat h$ becomes
\be
   \hat h =e^\Theta \,h \,e^{-\Theta}=
   -\frac{1}{2r_0^2\partial_z\bar\partial_{\bar
   z}\Theta}(\partial_z -2\partial_z \Theta)
   \bar\partial_{\bar z}
\ee

The generator of rotations about the origin in the projected plane
$\hat J=z\partial_z-\bar z\bar\partial_{\bar z}$ commutes with $\hat
h$. Then, we can diagonalize both simultaneously and decompose
$\hat\varphi$ as
\be
   \hat\varphi = z^j P(z\bar z).
\ee
By writing the undetermined function $P$ as a function of
\be
   x \equiv \frac{1-z\bar z}{1+z\bar z},
\ee
the eigenstate condition $\hat h \hat\varphi=E\hat\varphi$ leads
to the following ordinary differential equation for the function
$P(x)$,
\be
   (1-x^2)\frac{d^2 P}{dx^2} + (1-2j-3x)\frac{dP}{dx} + r_0^2 E P
   =0.
\ee

This is a differential equation of the Jacobi form, whose regular
solution in the interval [-1,1] is the Jacobi polynomial
\bea
   P_n^{(j,1-j)}(x) &\equiv& \frac{(-1)^n}{2^n n!}(1-x)^{-j}(1+x)^{j-1}
   \frac{d^n}{dx^n}\{(1-x)^{n+j}(1+x)^{n-j+1}\}\nonumber\\
   &=& \frac{1}{2^n} \sum_{m=0}^{n}
       \left( \begin{array}{c}
                    n+j\\
                     m
             \end{array}\right)
       \left( \begin{array}{c}
                    n-j+1\\
                     n-m
             \end{array}\right)(1-x)^{n-m}(1+x)^{m}
\eea
Note also that the Jacobi polynomial $P_n^{(j,1-j)}$ is defined
for $j\geq-n$, $1-j\geq-n$, this leading to upper and lower bounds
on the allowed values of $j$.

Putting all these together, we find that the orthonormal set of
eigenfunctions of $h=A\bar A$ results

\bea
    &&\varphi_{n}^j=\frac{1}{r_0}\sqrt{\frac{2n!(n+1)\Gamma(n+2)}{\pi\Gamma(n+j+1)
    \Gamma(n-j+2)}}\frac{z^j}{\sqrt{1+z\bar z}}\,P_n^{(j,1-j)}
    \left(\frac{1-z\bar z}{1+z\bar z}\right),\nonumber\\
    && E_n = \frac {1}{r_0^{2}}n(n+2),\quad\quad n=0,1,2,...\quad\quad
    j=-n, -n+1, ..., n+1
\eea

\section{Summary and discussion}

One of the goals of  this work was to construct and further analyze
a new type of string-like solutions in Chern-Simons theories which
is genuinely non-Abelian. To do this we adapted recent results
obtained in the four dimensional Yang-Mills theory context, where
non-Abelian strings, having many common features with QCD strings,
were founded in Yang-Mills theories. We started by constructing a
supersymmetric ${\cal N}=2$ Chern-Simons theory with $U(N)$ gauge
group and coupled to $N$ scalar multiplets. For this theory, we
found a Bogomol'nyi-like bound for the energy, this leading to a set
of first order self-duality equations. We solved the self-duality
equations by proposing a particular rotationally symmetric ansatz.
With this ansatz, equations for the profile functions reduced to
those of the Abelian Chern-Simons case. However, thanks to the
presence of the color-flavor rotations leaving the asymmetric vacuum
invariant, the solution proposed were found to have a non-trivial
orientational moduli space. By applying this color-flavor rotation,
we were able to write the gauge and Higgs fields in terms of the
orientational collective coordinates.

Another task we accomplished was to obtain the low energy theory
for the orientational moduli
space by assuming slowly-varying time-dependence for these
coordinates and then, substituting the resulting configuration in
the original Lagrangian. The resulting effective theory  corresponds
to a one dimensional ${\bf CP}^{N-1}$ sigma-model. Since self-dual
vortices are 1/2 BPS, the supersymmetric extension of the model
corresponds to  the  ${\cal N}=2$ ${\bf CP}^{N-1}$
sigma-model. We completed the analysis of the supersymmetric effective theory
by quantizing this low-energy theory  in the  $N=2$ case.

There are several aspects related to our results that
deserve future analysis. It particular, one should proceed to a
more complete analysis of the multi-vortex solutions. For this purpose,
it would be interesting to look for solutions similar to that
found in \cite{Auzzi:2005gr}, which consists in a composite state of two
coincident non-Abelian vortices in Yang-Mills theory. Another
relevant topic corresponds  to the case of semi-local non-Abelian vortices in
Chern-Simons theories,  expected to appear when the
number of flavor is larger than $N$.

Of particular interest is the study of Chern-Simons
non-Abelian vortices after quantization, since they could be
related to a new type of anyon-like object with possible
applications in planar systems. With this in mind, it would be
useful to further analyze at the quantum level the low-energy
theories for the orientational moduli. In particular, to
extend the $N=2$ results presented here to the
${\bf CP}^{N-1}$ model with generic $N$.

\section*{APPENDIX: Conventions for spinors in $d=2+1$}
We use the following representation for the $\gamma$ matrices,
\be
   (\gamma^0)_\alpha^{\;\beta} =
   (\sigma^2)_\alpha^{\;\beta},\quad
   (\gamma^1)_\alpha^{\;\beta}=i(\sigma^3)_\alpha^{\;\beta},\quad
   (\gamma^2)_\alpha^{\;\beta}=i(\sigma^1)_\alpha^{\;\beta},
   \label{repre}
\ee
where $\sigma$'s are the Pauli's matrices. Then, $\gamma$ matrices
satisfy
$$(\gamma^\mu)_\alpha^{\;\gamma}(\gamma^\nu)_\gamma^{\;\beta} = \eta^{\mu\nu}\delta_\alpha^{\;\beta}
 - i\epsilon^{\mu\nu\sigma}
(\gamma_\sigma)_\alpha^{\;\beta}.$$

The Lorentz rotation generator is realized in the following
manner:
\be
    {\cal M}^{\mu\nu}\equiv \frac i 4
    [\gamma^\mu,\gamma^\nu]=\frac 1 2
    \epsilon^{\mu\nu\rho}\gamma_\rho.
\ee
Using the representation (\ref{repre}) one can check the validity
of the following relation
\be
   \sigma^2{\cal M}^{\mu\nu}\sigma^2= - ({\cal M}^{\mu\nu})^T,
   \label{rel}
\ee
This gives as a result that $(\sigma^2 \psi)^T\chi $ is Lorentz
invariant. Therefore, we can define the spinor metric as
\be
    C^{\alpha\beta} \equiv i (\sigma^2)^{\alpha\beta} =
    \left(
          \begin{array}{cc}
          0 & 1 \\
          -1 & 0\\
          \end{array}
    \right) = C_{\alpha\beta},
\ee
which is used to raise and lower spinor indices via:
\be
   \theta_\alpha=\theta^\beta C_{\beta\alpha},\quad
   \theta^\alpha=C^{\alpha\beta}\theta_\beta
\ee

Note that $\gamma$ matrices are anti-Hermitian when lowering their
indices, i.e.,
\be
   (\gamma^0)_{\alpha\beta} = i\delta_{\alpha\beta},\quad
   (\gamma^1)_{\alpha\beta}=i(\sigma^1)_{\alpha\beta},\quad
   (\gamma^2)_{\alpha\beta}=-i(\sigma^3)_{\alpha\beta}.
\ee

Concerning the complex conjugation, we use barred spinors to
denote the complex conjugated of the unbarred ones, unlike the
usual definition using $\gamma^0$'s. Thus,
\be
   (\chi^\alpha)^*=\bar\chi^\alpha,\quad
   (\bar\chi^\alpha)^*=\chi^\alpha.
\ee
Besides, we define the conjugation of a product of spinors as
\be
   (\chi^\alpha\psi^\beta)^*=\bar\chi^\alpha\bar\psi^\beta
\ee

 \vspace{1 cm}

\noindent\underline{Acknowledgements}: L.A. wishes to thank the
Theory Division of CERN for hospitality while part of this work was
done. This work was partially supported by PIP6160-CONICET,
PIC-CNRS/CONICET, BID 1728OC/AR PICT20204-ANPCYT grants and by CIC
and UNLP.

\end{document}